\documentclass[conference]{IEEEtran}
\IEEEoverridecommandlockouts
\usepackage{cite}
\usepackage{amsmath,amssymb,amsfonts}
\usepackage{algorithmic}
\usepackage{graphicx}
\usepackage{textcomp}
\usepackage{xcolor}
\usepackage{subfigure}

\def\BibTeX{{\rm B\kern-.05em{\sc i\kern-.025em b}\kern-.08em
    T\kern-.1667em\lower.7ex\hbox{E}\kern-.125emX}}
\begin{document}

\title{Female mosquito detection by means of AI techniques inside release containers in the context of a Sterile Insect Technique program\\
\thanks{The participation of Javier Naranjo-Alcazar, Jordi Grau-Haro and Pedro Zuccarello in this research was funded by the Valencian Institute for Business Competitiveness (IVACE) and the FEDER funds by means of project MoTIA2 (IMDEEA/2022/70). The participation of David Almenar in this research was funded by Conselleria d’Agricultura, Desenvolupament Rural, Emergència Climàtica i Transició Ecològica of the Generalitat Valenciana and Subdirección de Innovación y Desarrollo de Servicios (grupo TRAGSA).}
}

\author{\IEEEauthorblockN{Javier Naranjo-Alcazar}
\IEEEauthorblockA{
\textit{Instituto Tecnológico de Informática (ITI)}\\
Valencia, Spain \\
jnaranjo@iti.es}
\and
\IEEEauthorblockN{Jordi Grau-Haro}
\IEEEauthorblockA{
\textit{Instituto Tecnológico de Informática (ITI)}\\
Valencia, Spain \\
jgrau@iti.es}
\and
\IEEEauthorblockN{David Almenar}
\IEEEauthorblockA{
\textit{Empresa de Transformación Agraria S.A., S.M.E., M.P. (TRAGSA)}\\
Paterna, Spain \\
dalmenar@tragsa.es}
\and
\IEEEauthorblockN{Pedro Zuccarello}
\IEEEauthorblockA{
\textit{Instituto Tecnológico de Informática (ITI)}\\
Valencia, Spain \\
pzuccarello@iti.es}
}

\maketitle

\begin{abstract}
The Sterile Insect Technique (SIT) is a biological pest control technique based on the release into the environment of sterile males of the insect species whose population is to be controlled. The entire SIT process involves mass-rearing within a biofactory, sorting of the specimens by sex, sterilization, and subsequent release of the sterile males into the environment. The reason for avoiding the release of female specimens is because, unlike males, females bite, with the subsequent risk of disease transmission. In the case of \emph{Aedes} mosquito biofactories for SIT, the key point of the whole process is sex separation. This process is nowadays performed by a combination of mechanical devices and AI-based vision systems. However, there is still a possibility of false negatives, so a last stage of verification is necessary before releasing them into the environment. It is known that the sound produced by the flapping of adult male mosquitoes is different from that produced by females, so this feature can be used to detect the presence of females in containers before environmental release. This paper presents a study for the detection of females in \emph{Aedes} mosquito release vessels for SIT programs. A particular set-up was developed to provoke the flight of mosquitoes inside the containers and record the sound produced by this flight. Due to the difficulty in identifying the exact timing of female flight within mixed male/female containers an anomaly detection strategy was chosen to be followed rather than a multi-class classification strategy. Two algorithms were implemented for the detection of female mosquitoes: an unsupervised outlier detection algorithm (iForest) and a one-class SVM trained with male-only recordings. The experiments conducted show that the results are highly dependent on the degree of maturity of the mosquitoes. For the detection of females with a 25/75\% female/male ratio, the best results are obtained on the seventh day after sexing in the pupal stage, reaching 87.5\% accuracy with iForest.
\end{abstract}


\section{Introduction}\label{sec:intro}

The so-called Sterile Insect Technique (SIT) \cite{dyck2021sterile} is a biological pest control technique, i.e. without the use of pesticides, based on the release into the environment of sterile male specimens of the species whose population is to be controlled. This technique began to be used on pests in the livestock environment. The reason for avoiding the release of females is because, unlike males, females sting, causing damage. Nowadays supported by the Joint Division of Nuclear Techniques in Food and Agriculture of the International Atomic Energy Agency (IAEA) and the Food and Agriculture Agency of the United Nations (FAO)\footnote{https://www.iaea.org/topics/sterile-insect-technique}, SIT is based on mass rearing of the insect in question inside a biofactory, separation by sex (or sex-sorting), sterilization of the males, and subsequent release of these into the environment.

In recent decades, SIT programs have begun to apply not only to agricultural pest control but also to insects that are potential vectors of contagious diseases for humans, such as \emph{Aedes} mosquitoes \cite{insects12030272, bellini2013pilot, balestrino2014validation}. In the case of this genre of mosquitoes, or others such as the \emph{Anopheles}, transmitters of malaria, avoiding the release of females into the SIT is crucial. Therefore, the sex-sorting stage in this type of pest control program is of vital importance and can be considered one of the enabling keys to the success of the program.

In the case of mosquitoes of the \emph{Aedes} family, sex-sorting for SIT programs has traditionally been performed by mechanical devices and systems that exploit the sexual dimorphism of the species in the pupal stage \cite{zacares2018exploring}, being the most accepted the Fay-Morlan device \cite{fay1959mechanical}. Some private sector initiatives have proposed systems based on machine vision, artificial intelligence and robotics to identify and separate adult female mosquitoes from male mosquitoes \cite{lepek2018method,desnoyer2020predictive, crawford2020efficient, zacares2018exploring}. However, as in every AI-based classification system, there is still a risk of false negatives in the sorting and separation system. Therefore, it makes sense to research and develop quality control systems that are capable of detecting the presence of adult females in the final stages prior to the release of sterile specimens. It is known that there are recognizable differences in the sound produced by the wing beat of male and female mosquitoes \cite{arthur2014mosquito},  therefore, it makes sense to take advantage of this phenomenon to detect the presence of female mosquitoes in the containers inside the biofactory in stages before the release of specimens into the environment.

In this work, the detection of female mosquitoes in containers where only male mosquitoes are expected to be present is considered. The presence of females is therefore considered an anomaly. This one-class approach was decided over a multi-class approach for biological reasons. It is very complex to record and label samples where both males and females are simultaneously flying when the ratio of females is very low. Although it is possible to produce canisters with the simultaneous presence of females and males, the flight of females is not assured. The main reasons for this are the size of the females (they are larger and therefore become tired earlier) and their attempt to remain unnoticed in an enclosed environment where they are surrounded by a large number of males. It should be noted that in a pot with a 50\%-50\% ratio of males to females, the wing beat of the females will be noticeable, but as the ratio of females decreases, their flight may appear only at specific and sparse moments in time that are difficult to identify for labeling.  Therefore, it is pointless to record 25\%-75\% male-female pots if the actual application has 25\% females or even less. The containers, a PVC tubular design of 8.8cm diameter and 12.5cm height \cite{insects13020178}, were placed in an experimental setup that allowed the recording of the sound of mosquito flight inside of them. Each container was filled with 250 specimens considering the cases of (i)~only male mosquitoes, (ii) only female mosquitoes, and (iii) 75\% males and 25\% females. Two algorithms were used to detect female gnats inside the containers from sound signal analysis: an unsupervised outlier detection algorithm (iForest) \cite{liu2008isolation} and a One-Class SVM \cite{scholkopf2001estimating} trained with male-only recordings. An audio-feature vector was extracted every 4s chunk by means of Google's TRILL model \cite{shor20_interspeech}.

As far as this group is aware, previous work in this field has focused on the detection of mosquitoes in the environment and/or on the classification among different species of them. The present work deals with a quality control stage inside a biofactory dedicated to a SIT program, therefore aiming to detect the presence of any female mosquitoes inside the canisters prior their release into the environment. To the authors' knowledge, this is the first time that computer listening has been applied to this problem in the described context.

\begin{figure}[t]
  \centering
  \centerline{\includegraphics[width=1.0\columnwidth]{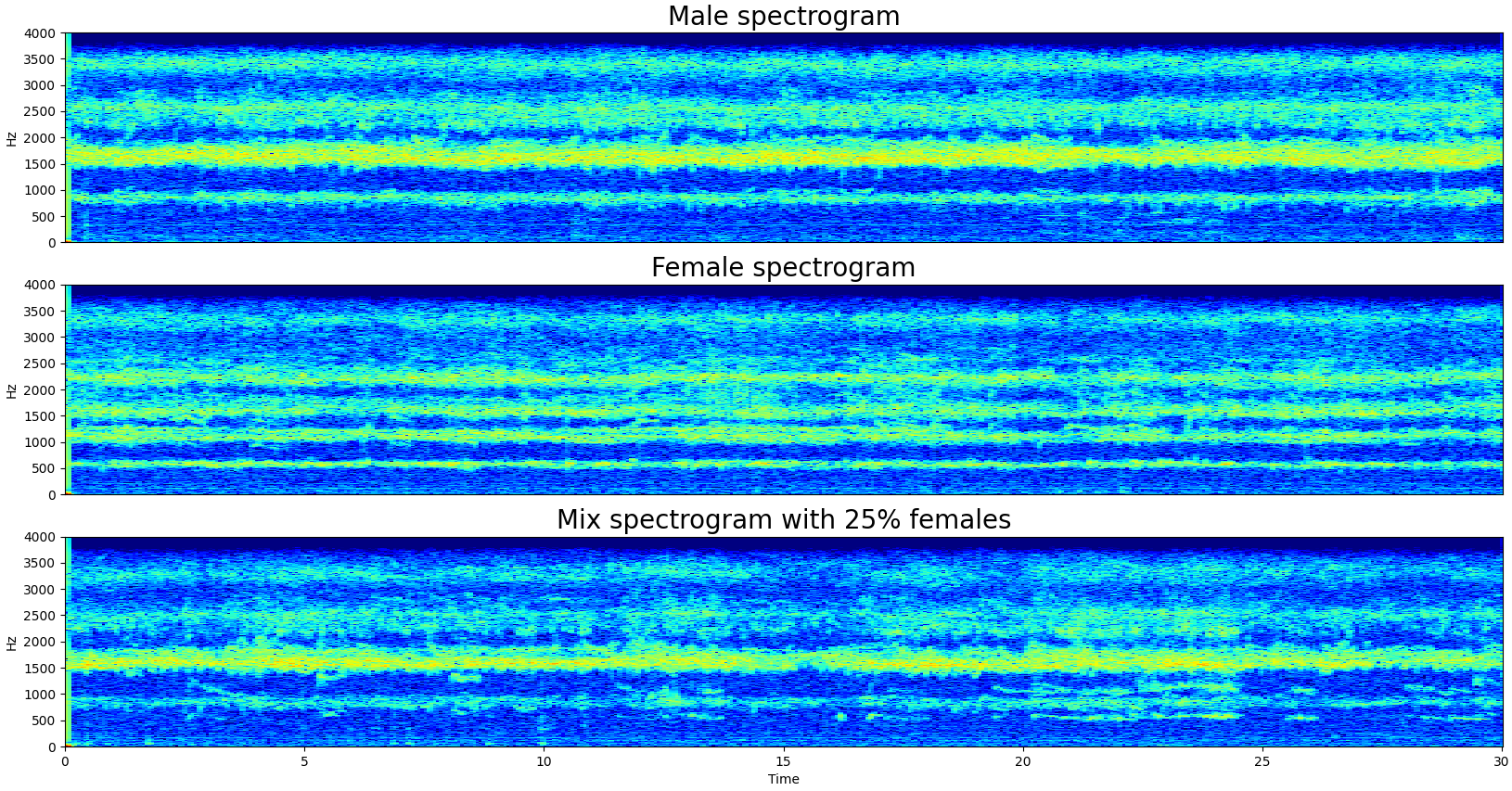}}
  \caption{Spectrogram of (top) 250 male mosquitoes, (middle) 250 female mosquitoes and (bottom) 250 mosquitoes with 75\% male and 25\% female ratios. Spectrogram of original audio signals subsampled at 8~kHz sampling rate.}
  \label{fig:all_specs}
\end{figure}

\section{Method}\label{sec:format}

\subsection{Audio processing}\label{subsec:audio}

The signals were captured with a sampling frequency of 44.1~kHz, however the frequencies at which the differences between the sound produced by males and females during the return are below 2~kHz. Therefore, the signals were filtered at 2~kHz and subsampled at 4~kHz.
Each signal was processed with the TRILL network \cite{shor20_interspeech} to obtain a feature vector. This network requires the sampling frequency of the input signal to be 16~kHz, so, in our case, the spectrogram of the signals would show null values in the range 2-8~kHz. This fact produces embeddings with negligible differences between the different scenarios to be discerned. Therefore, it would be desirable that only the frequency range of interest (below 2~kHz) serve as input to the TRILL network. This can be achieved by re-interpreting the sampling frequency of the signals that were subsampled at 4~kHz. By taking 4 seconds of these signals at 4~kHz and reinterpreting them as if they were 1 second sampled at 16~kHz, the desired zoom effect can be achieved. This reinterpretation of the sampling rate from 4~kHz to 16~kHz only affects the internal mapping of the signal when creating the log-Mel Spectrogram.



\begin{figure}[t!]
    \centering
    \subfigure[]{\includegraphics[width=0.48\columnwidth]{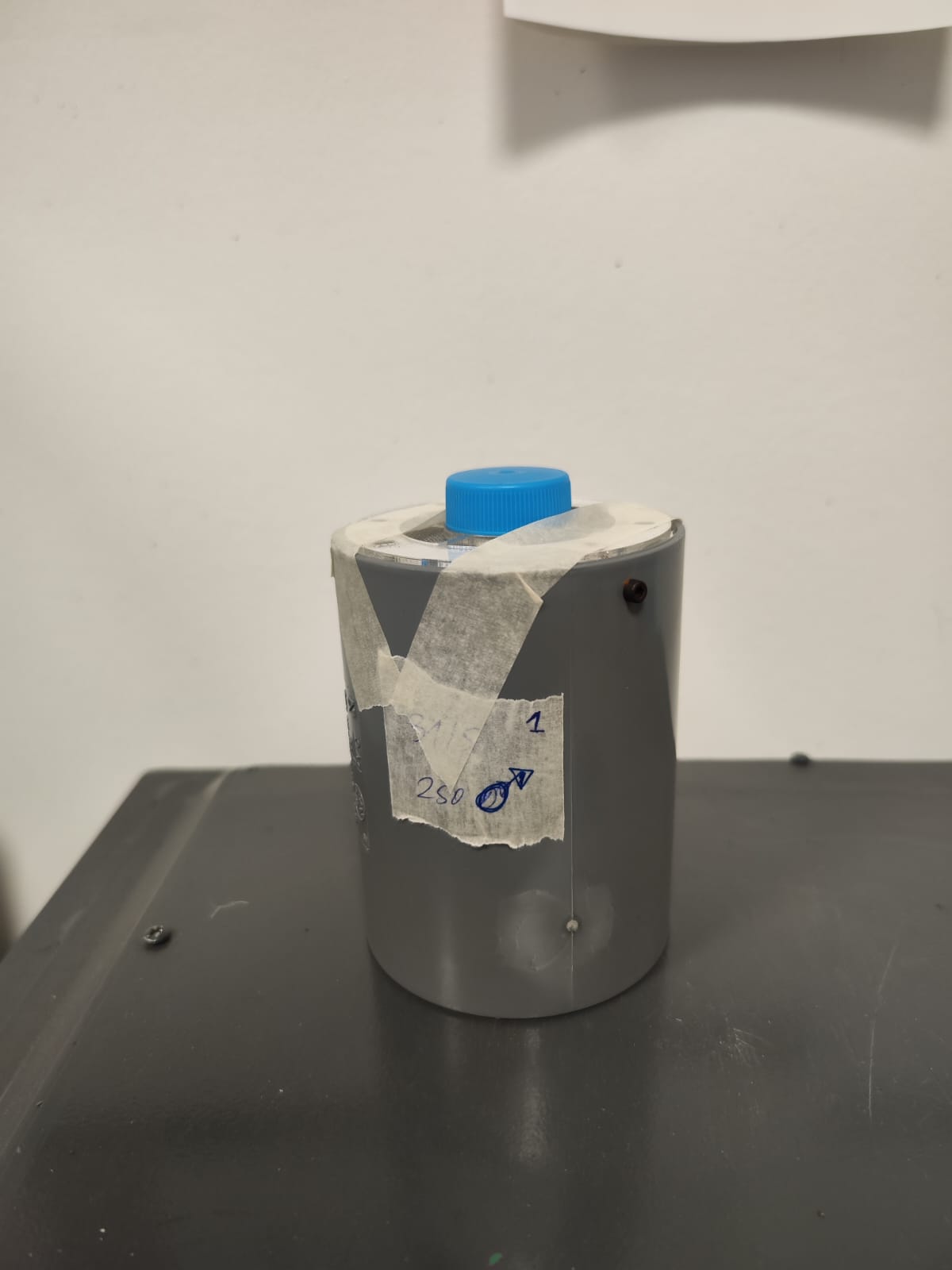}} 
    \subfigure[]{\includegraphics[width=0.48\columnwidth]{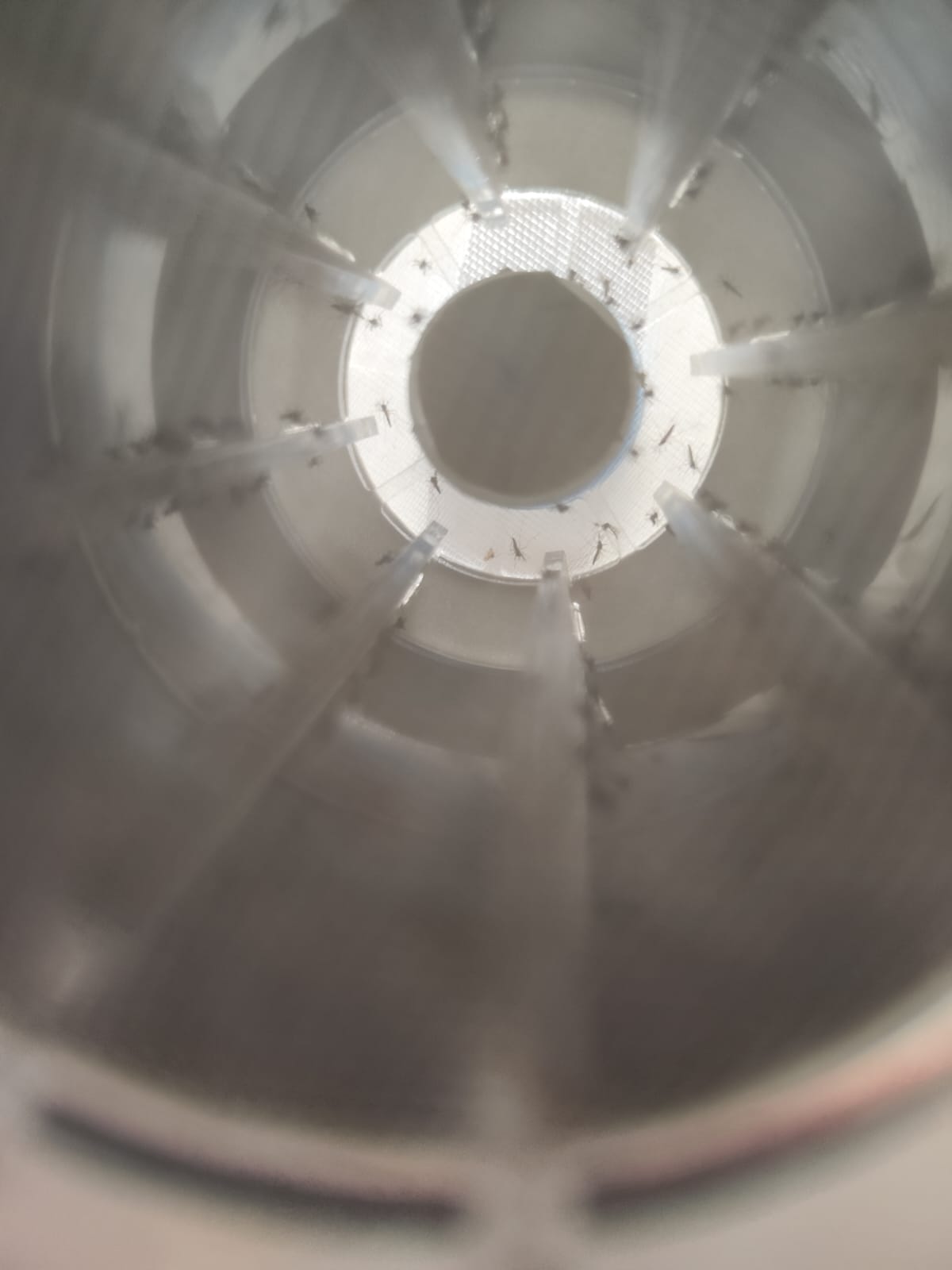}} 
    \subfigure[]{\includegraphics[width=0.48\columnwidth]{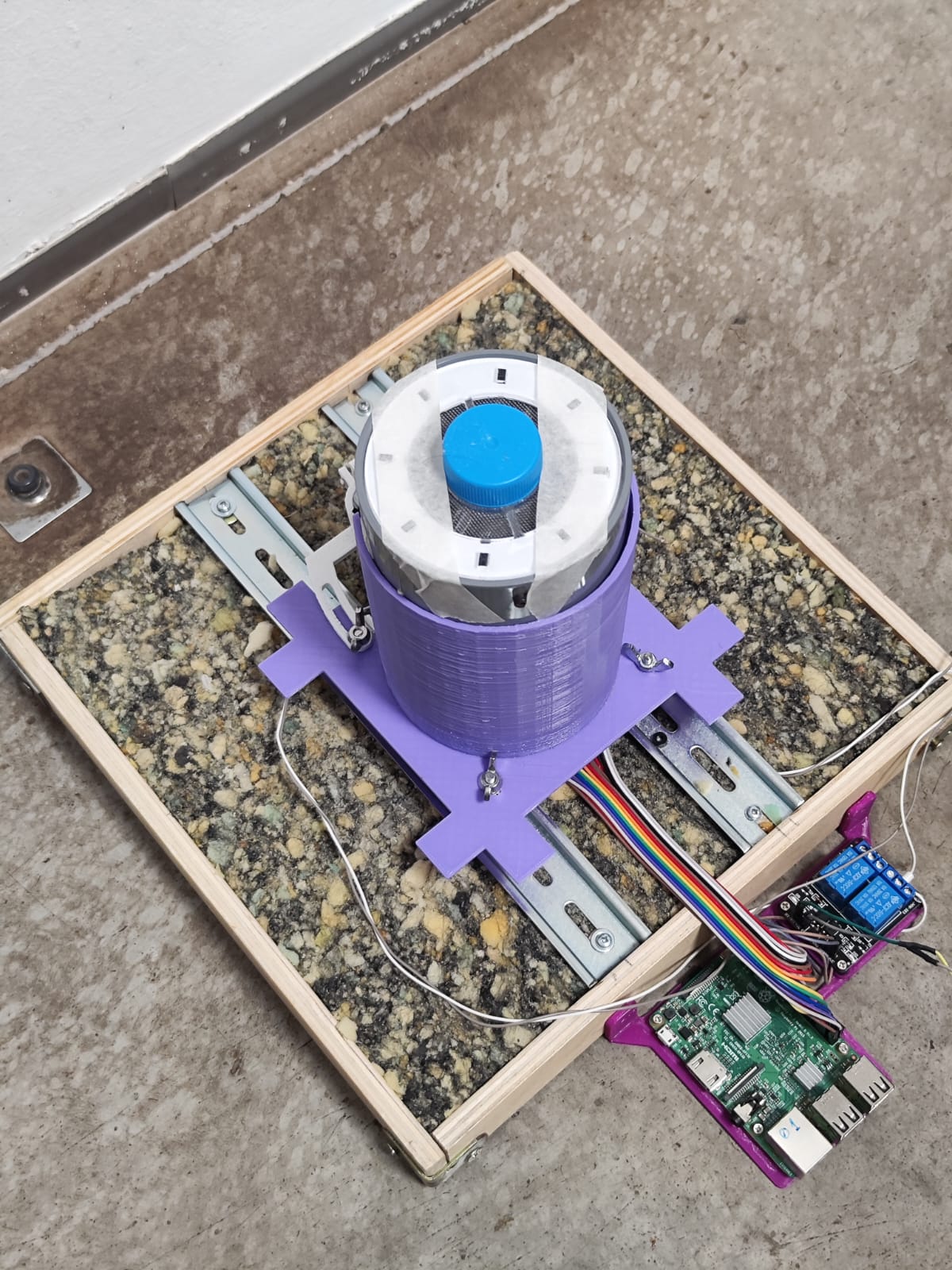}}
    \subfigure[]{\includegraphics[width=0.48\columnwidth]{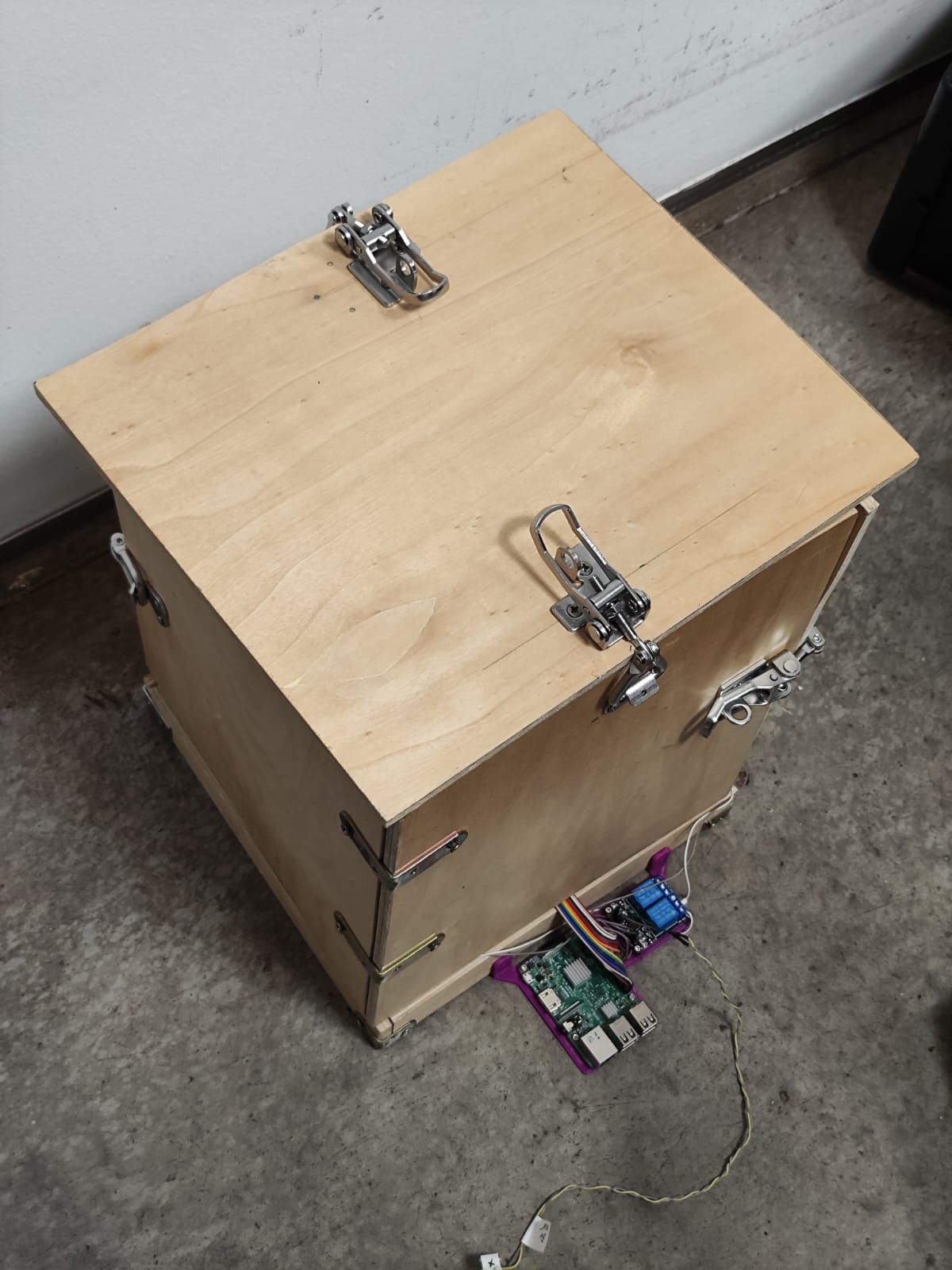}}
    \caption{(a) release container (b) inside container (c) open recording setup (d) recording setup}
    \label{fig:setup}
\end{figure}

A four-microphone array was used to capture sound at the bottom of the containers (see Fig.~\ref{fig:setup}). The audio signal used for processing and detection was the average of the four microphone signals.

\subsection{Outlier module}
\label{subsec:outlier}

Due to the number of samples produced in the experiment, it was decided to implement a solution based on transfer learning. The information from the mosquitoes' wing beat was found to exist in the range of 200 to 2000 Hz (see Fig.~\ref{fig:all_specs}). This frequency range is very similar to the human voice range. Therefore, it was decided to use a network trained with human voice segments, in our case TRIpLet Loss network \cite{shor20_interspeech} or TRILL for short. In earlier phases of the experiment, networks trained with environmental data were also tested\footnote{https://tfhub.dev/google/yamnet/1} \cite{kong2020panns}, but they did not perform well because the embeddings generated between the different containers were very similar. TRILL is a pre-trained network trained in a self-supervised manner with speech chunks from the Audioset dataset \cite{audioset} and released by Google through its TensorFlow Hub\footnote{https://tfhub.dev/google/nonsemantic-speech-benchmark/trill/3} platform. As explained in Sect.~\ref{subsec:audio}, the network was fed with 1-second chunks at 16kHz to obtain 512-position embedding.

Based on the difficulty of correctly tagging the behaviour of females in a container where males are also present, it was decided to design a system based on outlier detection. For this reason, an algorithm is trained with audios generated in only-males containers. In this way, the presence of the flight of a female could be considered an anomaly. The algorithms used for this study are the isolation Forest or iForest \cite{liu2008isolation} and the One-Class SVM (OCSVM) \cite{scholkopf2001estimating}. Both systems are trained with feature matrixes generated by the TRILL network with properly segmented audio-chanks (see Section~\ref{subsec:audio}). The library used for the implementation of the models is PyOD \cite{zhao2019pyod}. As all the training samples come from what is considered normal behavior, i.e. the flight of male mosquitoes, the contamination parameter required by the used algorithms is the lowest possible.

In the test stage, the 30-second audio clips are resampled as explained in Section~\ref{subsec:audio}. Subsequently, a prediction is made for each 4-second (@4~kHz) chunk with a 50\% overlap. Giving a total of 14 anomaly values for each audio clip (see Fig.~\ref{fig:plot_result}). If the mean of the anomaly scores is above a certain threshold, it is decided that the container is contaminated with female mosquitoes. In this work, a threshold equal to 0.5 has been considered.

\section{Experimental details}
\label{sec:exp_details}

\begin{figure}[t]
  \centering
  \centerline{\includegraphics[scale=0.35]{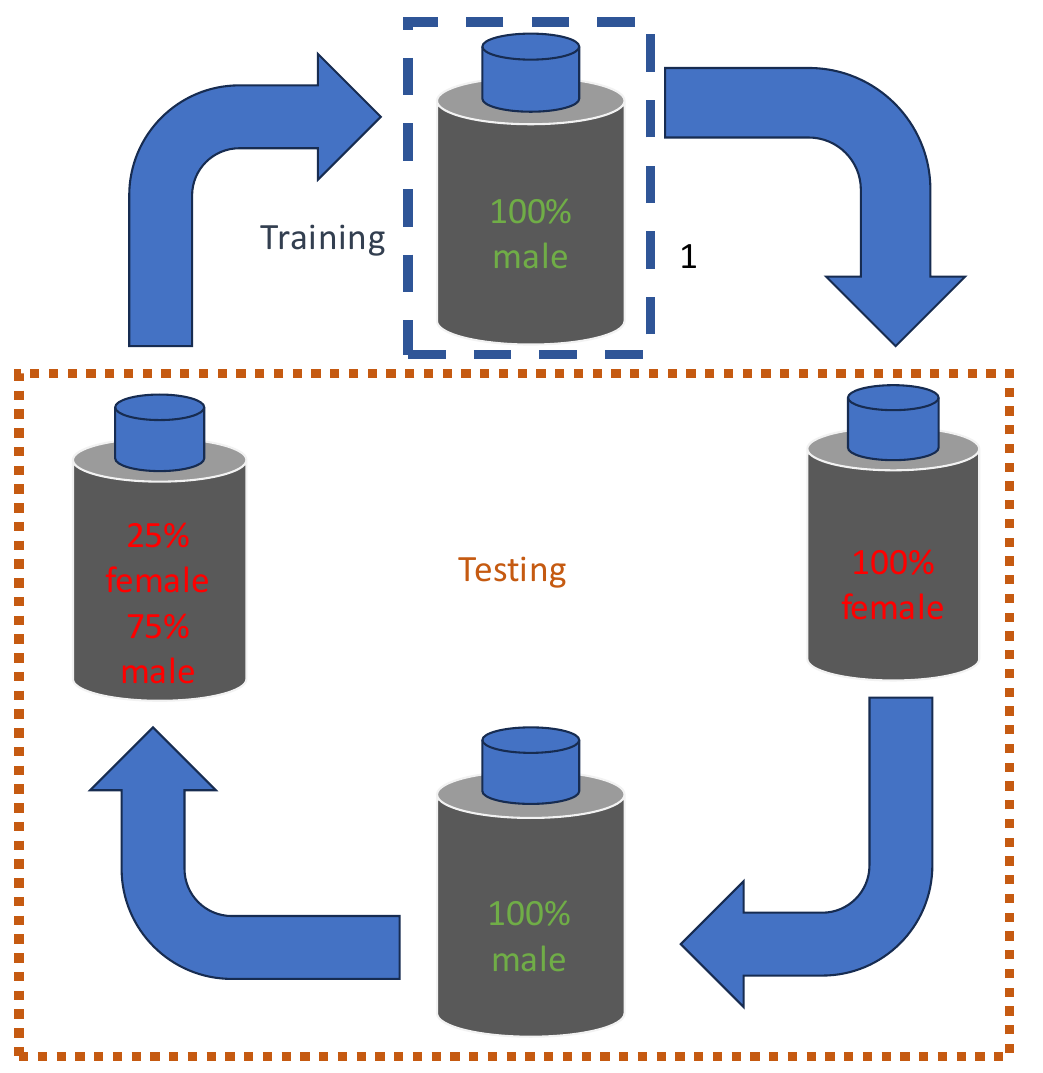}}
  \caption{Recording procedure carried out. This framework was carried out twice per day. In every interaction with a container, 8 audios of 30 seconds were recorded. The first recording container corresponds to the male container that have been used as a training set.}
  \label{fig:framework}
\end{figure}

\subsection{Recording setup}
\label{subsec:recording_setup}

The recording setup consists of 3 main components:

\begin{itemize}
    \item \underline{Release container:} consist of a two-piece tubular design where the external piece is a PVC pipe connection socket of 8.8~cm diameter and 12.5~cm height, while the internal piece consists of eight radial methacrylate strips interconnected by two rings that fit perfectly into the external piece. This container can be visualized in Fig.~\ref{fig:setup}-a). Each recipient is filled with 250 mosquitoes. Fig.~\ref{fig:setup}-b) shows several mosquitoes inside of one of the release canisters used in the experiment.
    \item \underline{Recording device:} the device used is the ReSpeaker 4-Mic Array\footnote{https://wiki.seeedstudio.com/ReSpeaker\_4\_Mic\_Array\_for\_Raspberry\_Pi/}  and it is is located under the container. The purpose of using an array of microphones is to obtain better spatial information since a mosquito may be flying from a particular side or corner. With this configuration, a greater spatial context of the mosquitoes' wing beat is obtained. The sampling rate during the recording process is 44.1~kHz. As explained in Section~\ref{subsec:audio} the signal is converted to mono by averaging the 4 channels. 
    \item \underline{Isolation box:} a wooden box has been made to isolate the sound of mosquitoes' wing beat. Fig.~\ref{fig:setup}-c) shows the inside of the box from any outside sound source. The release container is placed in the center of the wooden box and is held in place by a 3D-printed bracket. Underneath the container, the microphone array is placed. Fig.~\ref{fig:setup}-d) shows the final setup at the time of recording. The Raspberry Pi remains outside the box to avoid any noise produced by it. 
\end{itemize}

\subsection{Recording procedure}
\label{subsec:recording_procedure}

\begin{figure}[t]
  \centering
  \centerline{\includegraphics[width=\columnwidth]{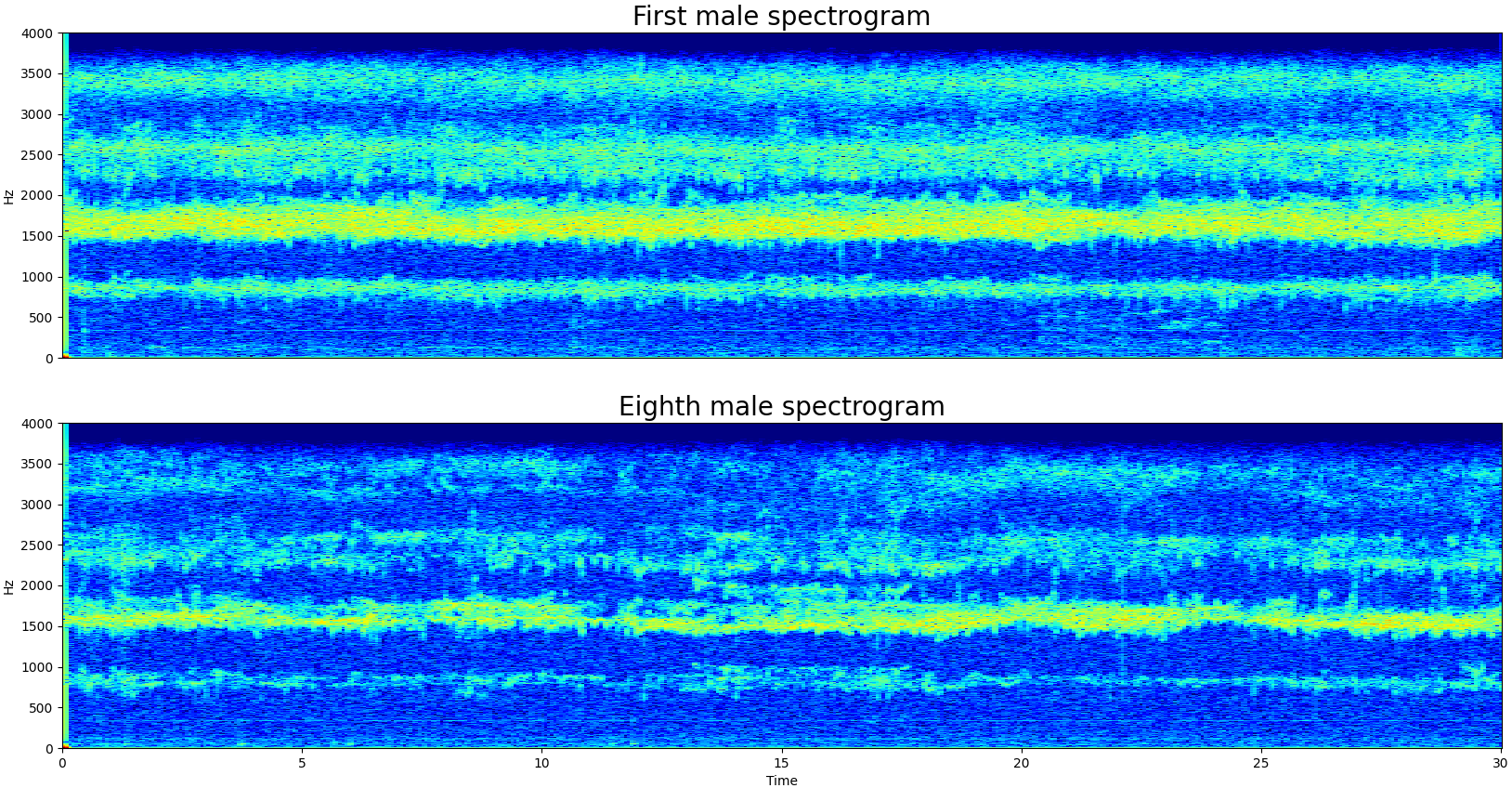}}
  \caption{Comparison of two spectrograms of the same container of males. The upper spectrogram corresponds to the first recorded clip (as soon as the container is introduced into the setup) and the second one once the container has been in the setup for 3.5 minutes.}
  \label{fig:male_specs}
\end{figure}

The recording process took place over four days (from 5 June to 8 June 2023)  at the facilities of \textit{Conselleria d’Agricultura, Desenvolupament Rural, Emergència Climàtica i Transició Ecològica of the Generalitat Valenciana}. The mosquitoes were reared and sex-sorted at pupal stage according to the protocol described in \cite{insects12030272}. Thus, the first day of recording corresponds to the 6th day after sex-sorting. The recording took place each day at 14:30 hrs. 
The order of the containers during the recording was as follows: container of males (training set), container of females, container of males (for test) and mix container (25\% female and 75\% male).  From each container, 8 audios of 30 seconds duration were recorded. The recording cycle was performed twice, producing a total of 16 30-second recordings per recipient per day. Two batches of 8 clips are recorded instead of 16 at once to avoid the mosquitoes getting used to the setup. It was empirically proven that after 4 minutes inside the setup, the mosquitoes became habituated and flew much less. This fact can be seen in Fig.~\ref{fig:male_specs}, the frequency components are much attenuated in the spectrogram in the image below. For a better understanding, Fig.~\ref{fig:framework} shows an example of the daily recording process.

\subsection{Dataset}
\label{subsec:dataset}

The dataset used for this study consists of 2 hours and 8 minutes of recording. The recording of each container corresponds to a total of 32 minutes. Thus, there are 8 minutes of recording per container per day. It should be noted that the training sessions are independent among days as the mosquitoes may vary in their behaviour. This occurrence is visible in Table~\ref{tab:results}. 

Each system has accordingly been trained using part of the 8 minutes available in the training container (16 clips of 30 seconds). From each clip, 6 segments of 4 seconds are manually segmented without overlap. Therefore, the training set consists of 96 clips of 4 seconds duration.

\begin{figure}[t!]
    \centering
    \subfigure[]{\includegraphics[width=\columnwidth]{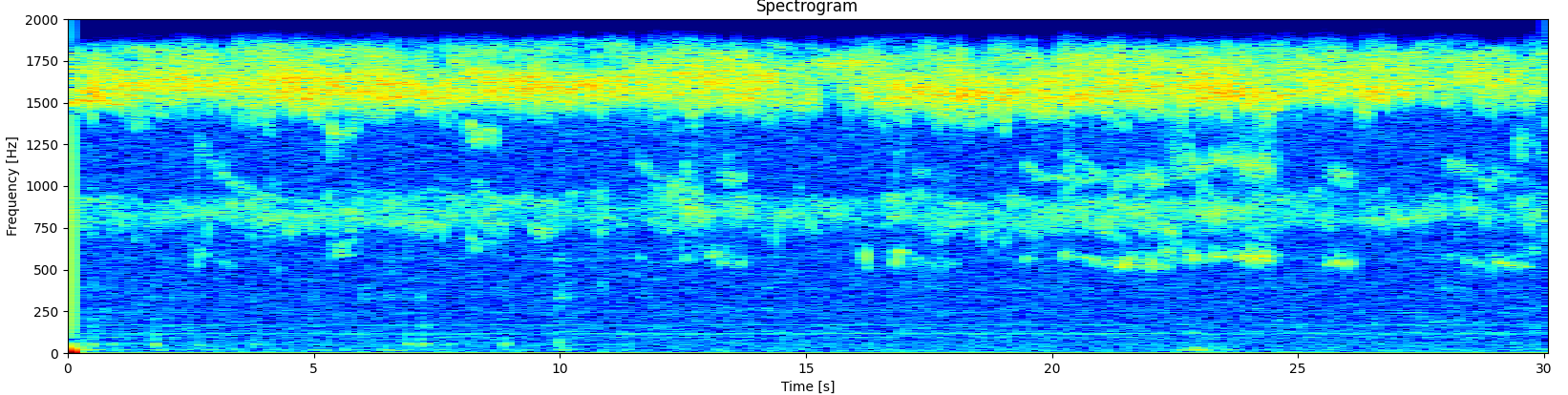}} 
    \subfigure[]{\includegraphics[width=\columnwidth]{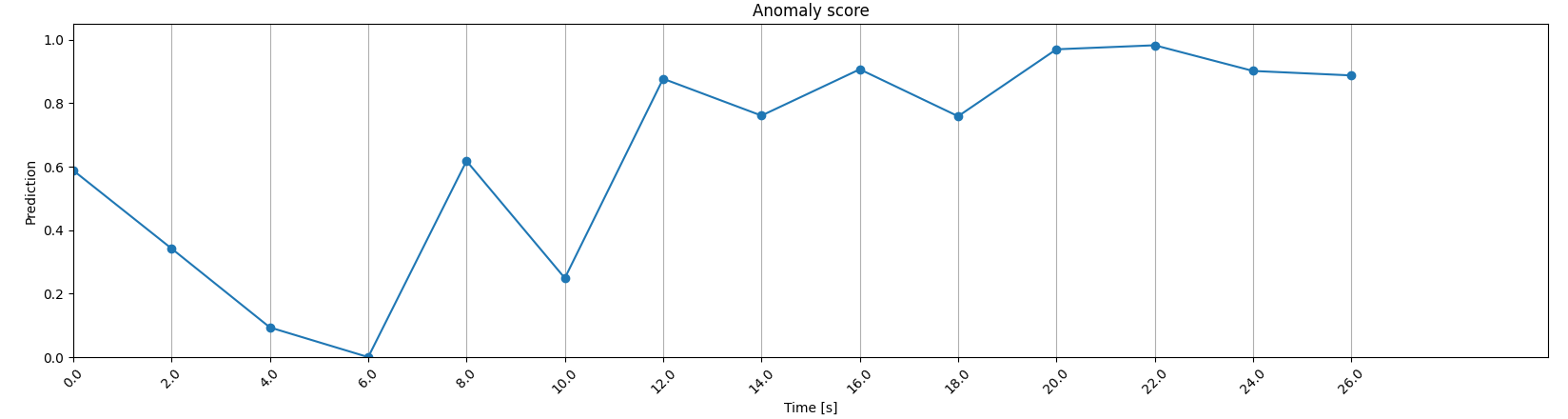}}
    \subfigure[]{\includegraphics[width=\columnwidth]{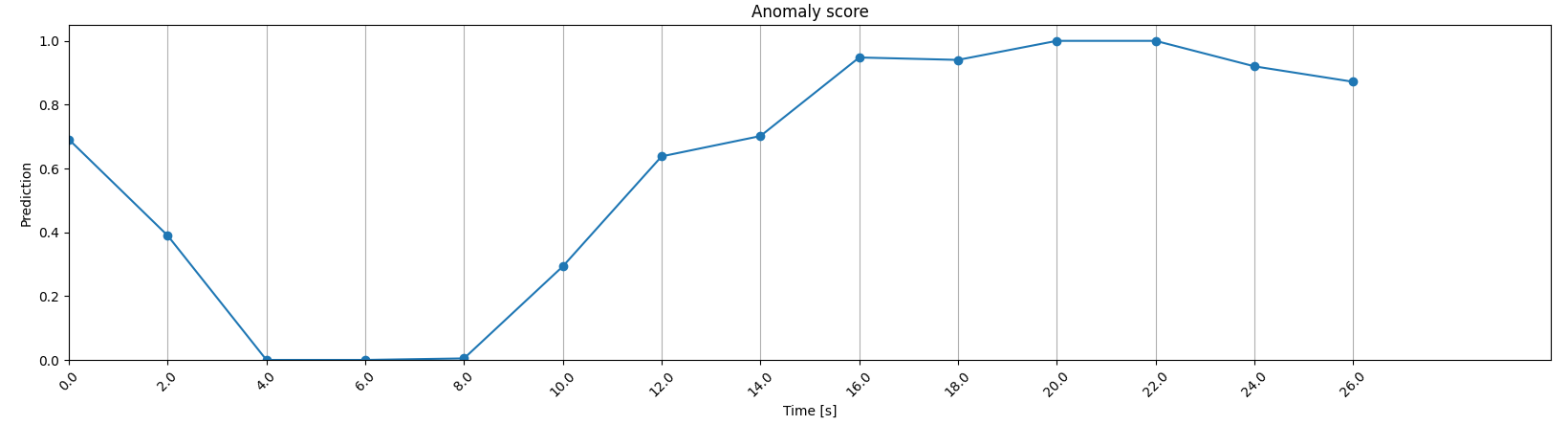}} 
    \caption{(a) Spectrogram of a 25\%/75\% female/male container audio of the 6th day since sex-sorting (b) iForest system output (c) OCSVM system output. The prediction point at $t=n$ seconds is the prediction for $t=n$ seconds to $t=n+4$ seconds audio segment.}
    \label{fig:plot_result}
\end{figure}

\begin{table}[]
\centering
\begin{tabular}{c|cc|cc|}
\hline
\multicolumn{1}{|c|}{Day} & \multicolumn{2}{c|}{100\% male}            & \multicolumn{2}{c|}{25\% female 75\% male}            \\ \hline
                          & \multicolumn{1}{c|}{OCSVM} & iForest        & \multicolumn{1}{c|}{OCSVM} & iForest        \\ \hline
\multicolumn{1}{|c|}{6th} & \multicolumn{1}{c|}{75}    & 81.25          & \multicolumn{1}{c|}{68.75} & 75             \\ \hline
\multicolumn{1}{|c|}{7th} & \multicolumn{1}{c|}{93.75} & \textbf{81.25} & \multicolumn{1}{c|}{75}    & \textbf{87.5}      \\ \hline
\multicolumn{1}{|c|}{8th} & \multicolumn{1}{c|}{56.25} & 56.25          & \multicolumn{1}{c|}{62.5}  & 68.75              \\ \hline
\multicolumn{1}{|c|}{9th} & \multicolumn{1}{c|}{85.7}  & 75             & \multicolumn{1}{c|}{56.25} & 62.5              \\ \hline
\multicolumn{1}{|c|}{All} & \multicolumn{1}{c|}{73.40} & \textbf{73.43} & \multicolumn{1}{c|}{65.62} & \textbf{73.43}     \\ \hline
\end{tabular}
\caption{Accuracy on the different test bottles used in the experiment. The column ``Day" indicates the day since the pupae were sexed. The case of 100\% females is not shown because its result is trivial.}
\label{tab:results}
\end{table}

\section{Results}
\label{sec:results}

The daily results as well as the average results per test recipient can be seen in Table~\ref{tab:results}. First of all, as expected, both solutions are able to detect as anomalous sounds the containers containing only females (100\% females). In this case, the day does not influence the performance of the system. This trivial study has been carried out to make a small analysis of female wing beat that would allow a better comparison with the mixed containers.

A considerable fact to note is the change in the performance of the system depending on the day. As can be seen, there is a considerable improvement in both systems between the sixth and seventh day. In particular, the system using the iForest achieves 87.5\% accuracy in detecting females in the 25\%/75\% female/male container while obtaining 81\% accuracy in the 100\% male. On the other hand, both systems drop in performance on the eighth day in both, the 100\% male and 25\%/75\% female/male containers. In contrast, on the ninth day, there was an improvement in accuracy in the 100\% male recipient, but still a worsening performance in detecting females. This could be due to multiple factors such as female fatigue (females have a larger size and get tired earlier than males) or comfort in the container. These results indicate that the seventh day after sex-sorting seems to be the optimal day for detecting female mosquitoes.

Fig.~\ref{fig:plot_result} corresponds to the outputs of the two systems on the same audio. The audio corresponds to the 25\%/75\% female/male container on the first day of recording (6th day since sex-sorting). As can be seen, the iForest algorithm (Fig.~\ref{fig:plot_result}-b)) is much more sensitive to the appearance of the female. At instant $t=8$ the anomaly score of the OCSVM is 0, indicating that the algorithm detects that chunk, from 8 to 12 seconds, as a male wing beat. In contrast, the iForest shows an anomaly score of about 0.6. In the spectrogram, a small anomaly can be observed at that instant in time.

To summarise, it can be concluded that the day of recording affects the performance of the system, as the fluttering of the females is essential for their detection. It can be appreciated that the seventh day after sex-sorting seems to be the better day for female-detection. With regard to the systems analyzed, the iForest system shows a greater sensitivity to the appearance of anomalies, as well as showing a good trade-off between the detection of females and the characterisation of the male wing beat.

\section{Conclusions}
\label{sec:conclusion}

An efficient and accurate sex-sorting technique for mosquitoes is of vital importance for SIT programs. Today, these task is performed using mechanical devices manually operated by expert workers, although in very few cases some AI-based vision systems are being implemented.

This paper proposes a quality control strategy for the detection of false negatives in Aedes mosquito sexing systems. The approach is based on the analysis, by means of AI, of the sound produced by the flight of the mosquitoes inside the containers that would later be used to release them into the environment. The proposal is based on the contracted fact that the flight of male and female mosquitoes produces sufficiently distinguishable sounds. The normal situation is that the containers contain only male mosquitoes, therefore, in our system, the presence of females is considered an anomaly from a mathematical point of view.  Following this idea, two sound anomaly detection algorithms have been tested, iForest and OCSVM.

Some audio processing is performed to enhance the relevant frequency band where the information from the mosquitoes' wing beat resides. Experiments were carried out using containers with (i) 100\% male mosquitoes, (ii) 100\% females and (iii) 25\%/75\% females/males. In all cases there were 250 mosquitoes. Due to the changing behaviour of the insects depending on the day since the sex-sorting, the experiments were conducted on four consecutive days. The solution based on the iForest algorithm shows promising results if used on the seventh day after sex-sorting in the pupal stage. An 87.5\% accuracy in detecting females in a container consisting of 25\%/75\% females/males was obtained. If averaged over all days, the accuracy is 73\%. On the other hand, the detection of male-only recipients (normal behaviour) is also at its best on the seventh day with 81\%. Its average is also 73\%.


\bibliographystyle{IEEEtran}
\bibliography{refs}

\end{document}